\documentclass[aps,prd,nofootinbib,twocolumn,superscriptaddress,showpacs,floatfix]{revtex4}
\usepackage{mathrsfs}
\usepackage{latexsym}
\usepackage{amsmath}
\usepackage{amssymb}
\usepackage{graphicx}
\usepackage{longtable}
\usepackage{multirow}
\usepackage{bbm}
\usepackage{epsfig}
\usepackage[usenames]{color}
\usepackage[colorlinks=true,citecolor=blue,linkcolor=blue]{hyperref}
\allowdisplaybreaks

\begin{document}



\title{Deformed QCD phase structure and entropy oscillation in the presence of a magnetic background}

\author{Guo-yun Shao}
\thanks{These authors contributed equally to this work}
\affiliation{School of Science, Xi'an Jiaotong University, Xi'an, 710049, China}
\affiliation{ MOE Key Laboratory for Nonequilibrium Synthesis and Modulation of Condensed Matter, Xi'an Jiaotong University, Xi'an, 710049, China}

\author{Wei-bo He} 
\thanks{These authors contributed equally to this work}
\affiliation{School of Science, Xi'an Jiaotong University, Xi'an, 710049, China}

\author{Xue-yan Gao}
\affiliation{School of Science, Xi'an Jiaotong University, Xi'an, 710049, China}

\begin{abstract}

The QCD phase transitions are investigated in the presence of an external magnetic field in the Polyakov improved Nambu--Jona-Lasinio (PNJL)  model. 
We detailedly analyze that how the filling of multiple Landau levels by light~(up and down) quarks deforms the QCD phase structure under different magnetic fields.
In particular, we concentrate on the phase transition under a magnetic field possibly reachable in the non-central heavy-ion collisions at RHIC.  
The numerical result shows that  two first-order transitions or more complicate phase transition in the light quark sector can exist for some magnetic fields, different from the phase structure under a very strong or zero magnetic field. These phenomena are very interesting and possibly relevant to the non-central heavy-ion collision experiments with colliding energies at several $A$ GeV as well as the equation of state of magnetars.  Besides, we investigate the entropy oscillation with the increase of baryon density in a magnetic background.

\end{abstract}


\maketitle

\section{Introduction}
Over the decades, intensive investigations have been performed to explore the structure of  strongly interacting matter. 
At high temperature and small chemical potential, the heavy-ion collision experiments indicate that the transformation from quark-gluon plasma~(QGP) to hadrons  is a smooth crossover~\cite{Gupta11}, which is consistent with the lattice QCD (LQCD) calculations~\cite{Aoki06, Bazavov12, Borsanyi13,Bazavov14, Bazavov17,Borsanyi14}.   A first-order phase transition, with a critical endpoint (CEP) connecting with the crossover transition, is predicted at large chemical potentials by some popular quark models which incorporate the symmetry of QCD~(e.g.,\cite{Fukushima04,Ratti06,Costa10,Fu08, Sasaki12, Schaefer10, Skokov11, Qin11,Gao16, Fischer14,Shi14}).  Searching for the critical endpoint is one of the primary tasks of RHIC STAR~\cite{Aggarwal10, Adamczyk14}. The second phase of the beam energy scan~(BES-II) at STAR is being performed with enhanced luminosity, focusing on the energy range  $\sqrt{s_{_{NN}}}=7.7\sim20$\,GeV where some possible indications related to critical phenomenon were reported  based on the preliminary result of BES-I~\cite{Luo2014, Luo2016, Luo2017}.

A more challenging question is how the properties of strongly interacting matter will be affected when an external magnetic field emerges~(for recent reviews, please refer to \cite{Kharzeev13, Miransky15, Aadersen16} ).  There are at least two areas related to strong interaction where magnetic field plays a very important role: magnetars and non-central heavy-ion collisions.
In the core of magnetars, the magnetic field strength possibly reach $10^{18}\sim10^{20}\,$G~\cite{Bocquet95,Ferrer10}, which gives birth to a stiffer equation of state of neutron star matter and thus can support a massive compact star. 
In the non-central heavy-ion collisions, the intensity of magnetic field depends on the beam energy and centrality.  The magnetic field, up to $B=10^{19}\,$G or $eB\sim 6m_{\pi}^2$, is possibly created at RHIC~\cite{Kharzeev08,Skokov09}, while up to $eB\sim 15m_{\pi}^2$ can possibly be reached at LHC~\cite{Voronyuk11,Bzdak12,Deng12}.  In all the above situations, the magnetic field intensity can be the same order as or larger than $\Lambda_{QCD}$ scale, therefore it will definitely produce a profound effect on the QCD phase transition.

There are two main aspects in the study of strongly interacting matter under an external magnetic field related to heavy-ion collisions. One is the chiral magnetic effect~(CME)~\cite{Kharzeev08,Buividovich09} and the related phenomena such as the chiral separation effect~(CSE)~\cite{Kharzeev11} and the chiral magnetic wave~(CMW)~\cite{Metlitski05, Burnier11, Gorbar11}. The essence of the CME is the imbalance of the chirality. The possibility that the CME can be observed in heavy-ion collisions has stimulated the exploration of strong interactions in presence of a chirality imbalance and a magnetic field~\cite{Fukushima08,Fukushima10,Gatto12,Ruggieri11,Fu2013,Bayona11}. 

The other aspect is the QCD phase transition driven by a strong magnetic field. At zero temperature, the lattice studies indicate that the chiral condensate tends to increase with the increasing magnetic field, which is knowns as magnetic catalysis~(MC)~\cite{Buividovich10}. It means that the magnetic field contributes to the chiral symmetry breaking. Correspondingly, it suggests that the critical temperature of chiral symmetry restoration should be enhanced for a stronger magnetic field. Later lattice calculation with the pion mass in the range $200-480\,$MeV indeed shows that the critical temperature slightly increases with the enhancement of magnetic field~\cite{elia, Ilgenfritz12}. However,  when the physical pion mass $m_{\pi}=145\,$MeV is taken, the inverse magnetic catalysis~(IMC) occurs, i.e., the increase of magnetic field tends to suppress the quark condensate near the critical temperature of chiral transition, and lowers the phase transition temperature at zero chemical potential~\cite{Bali:2011qj,Bali:2012zg}. 
The discovery of IMC effect at high temperature has motivated the improvements of the effective quark models to give a consistent result with LQCD calcualtion. Different mechanisms have been proposed in literatures to explore the IMC effect and QCD phase transitions~( see, e.g., \cite{Fukushima12,Chao13,Fraga13,Fukushima13,Fukushima16,Fayazbakhsh14,Costa15,Ferreira142,Fraga14,Yu15,Ayala16,Pagura17,Menezes:2008qt}).

In the present study, we are more interested in the complete QCD phase diagram under a background magnetic field.  
Related studies with relatively larger magnetic fields have be done in Ref.~\cite{Ferreira18} and meaningful results about the phase transition in strange quark sector have been achieved. 
However, the phase diagram at low temperatures and densities where up and down quarks dominate strongly depends on the magnetic field intensity. How the QCD phase structure changes from small to large magnetic fields needs to be explored.
We herein will focus on the  QCD phase structure under relatively smaller magnetic field.
In particular, we will detailedly analyze the relation between the deformed first-order transition and the filling of Landau levels by up and down quarks.   In addition,
we will study the entropy oscillation with the increasing baryon density under an external magnetic field.
The study is in some degree relevant to the non-central collisions at STAR  where relatively smaller magnetic field can be generated.%
%
%

The paper is organized as follows. In Sec.~II, we introduce the thermodynamics of quark matter in the presence of an external magnetic field within the 2+1 flavor PNJL quark model. In Sec.~III, we illustrate the numerical results of the deformed QCD phase structure under different magnetic fields, and discuss the influence of  the filling of Landau levels on the phase transition as well as the entropy oscillation along baryon density. A summary is finally given in Sec. IV.

\section{ Thermodynamics of magnetized quark matter}

We first briefly introduce the thermodynamics of magnetized quark matter in the 2+1 flavor PNJL model. In the presence of an external magnetic field, the Lagrangian density takes the form, 
\begin{eqnarray}
\mathcal{L}&\!=&\!\bar{q}(i\gamma^{\mu}D_{\mu}\!+\!\gamma_0\hat{\mu}\!-\!\hat{m}_{0})q\!+\!
G\sum_{k=0}^{8}\big[(\bar{q}\lambda_{k}q)^{2}\!+\!
(\bar{q}i\gamma_{5}\lambda_{k}q)^{2}\big]\nonumber \\
           &&-K\big[\texttt{det}_{f}(\bar{q}(1+\gamma_{5})q)+\texttt{det}_{f}
(\bar{q}(1-\gamma_{5})q)\big]\nonumber \\ \nonumber 
&&-U(\Phi[A],\bar{\Phi}[A],T)-\frac{1}{4}F^{\mu\nu}F_{\mu\nu},
\end{eqnarray}
where $q$ denotes the quark fields with three flavors, $u,\ d$, and
$s$; the current mass $\hat{m}_{0}=\texttt{diag}(m_{u},\ m_{d},\
m_{s})$ and the quark chemical potential $\hat{\mu}=\texttt{diag}(\mu_u,\mu_d,\mu_s)$ in flavor space; $G$ and $K$ are the four-point and
six-point interacting constants, respectively.  
The covariant derivative is defined as $D^\mu=\partial^\mu-iA^\mu-iq_iA_{EM}^{\mu}$, where $ A_\mu =g\mathcal{A}_{\mu}^a\frac{\lambda_a}{2}$, in which $\mathcal{A}_{\mu}^a$ represents the SU(3) gauge field and $\lambda_a$ are the Gell-Mann
matrices.  $A_{EM}^{\mu}$ is the electromagnetic vector potential, and  $A_{EM}^{\mu}=\delta^{\mu2}x_1B$ for a  static and constant magnetic field in the $z$ direction.
$F^{\mu\nu}=\partial^{\mu}A_{EM}^{\nu}-\partial^{\nu}A_{EM}^{\mu}$ are used to account for the external magnetic field.

The effective potential $U(\Phi[A],\bar{\Phi}[A],T)$ is expressed with the traced Polyakov loop
$\Phi=(\mathrm{Tr}_c L)/N_C$ and its conjugate
$\bar{\Phi}=(\mathrm{Tr}_c L^\dag)/N_C$. The Polyakov loop $L$  is a matrix in color space
\begin{equation}
   L(\vec{x})=\mathcal{P} exp\bigg[i\int_0^\beta d\tau A_4 (\vec{x},\tau)   \bigg],
\end{equation}
where $\beta=1/T$ is the inverse of temperature and $A_4=iA_0$.
The Polyakov-loop effective potential in present study takes the form
%
\begin{eqnarray}
     \frac{U(\Phi,\bar{\Phi},T)}{T^4}&=&-\frac{a(T)}{2}\bar{\Phi}\Phi +b(T)\mathrm{ln}\big[1-6\bar{\Phi}\Phi\\ \nonumber
                                                &&+4(\bar{\Phi}^3+\Phi^3)-3(\bar{\Phi}\Phi)^2\big],
\end{eqnarray}
where
$a(T)=a_0+a_1(\frac{T_0}{T})+a_2(\frac{T_0}{T})^2$ and $ b(T)=b_3(\frac{T_0}{T})^3$.
The parameters $a_0=3.51$, $a_1=-2.47$,  $a_3=15.2$, and $b_3=-1.75$ were derived in~\cite{Robner07} by fitting the thermodynamics of pure gauge sector in LQCD.
$T_0=210$\, MeV is implemented when fermion fields are included
%
%

 
 In the mean field approximation, the thermodynamical potential of magnetized quark matter can be derived as~\cite{Fu2013}
 \begin{eqnarray}
 \Omega&=&\sum_{f=u, d, s}(\Omega_{f}^{\mathrm{0}}+\Omega_{f}^{\mathrm{T}} )+2G(\phi_u^{2}+\phi_d^{2}+\phi_s^{2}) \nonumber \\
 &&+4 K\phi_u\phi_d\phi_s+U(\Phi, \overline{\Phi}, T),
\end{eqnarray}
where
 \begin{eqnarray}
\begin{aligned} \Omega_{f}^{\mathrm{0}}=-N_{c}\frac{\left|q_{f}\right| e B}{2 \pi} \sum_{n=0}^{\infty} \alpha_{n} \int_{-\infty}^{\infty} \frac{d p_{z}}{2 \pi} E_{f,n}\end{aligned},
\end{eqnarray}
and
 \begin{eqnarray}\label{5}
\begin{aligned} \Omega_{f}^{\mathrm{T}}=-T \frac{\left|q_{f}\right| e B}{2 \pi} \sum_{n=0}^{\infty} \alpha_{n} \int_{-\infty}^{+\infty} \frac{d p_{z}}{2 \pi}\left(Z_{f}^{+}+Z_{f}^{-}\right)\end{aligned}.
\end{eqnarray}
In Eq. \ref{5}, 
 \begin{equation}
Z_f^{+}\!=\!\mathrm{ln}(1\!\!+\!\!3\Phi e\!^{-\frac{E_{f,n}-\mu_f}{T}}\!\!+\!\!3\bar{\Phi}e^{-2\frac{E_{f,n}-\mu_f}{T}}\!\!+\!\!e^{-3\frac{E_{f,n}-\mu_f}{T}}),
\end{equation}
and
  \begin{equation}\label{}
Z_f^{-}\!=\!\mathrm{ln}(1\!\!+\!\!3\bar\Phi e\!^{-\frac{E_{f,n}+\mu_f}{T}}\!\!+\!\!3{\Phi}e^{-2\frac{E_{f,n}+\mu_f}{T}}\!\!+\!\!e^{-3\frac{E_{f,n}+\mu_f}{T}}).
\end{equation}
The quasi-particle energy is
\begin{equation}\label{Ein}
E_{f,n}=(2n|q_f|B+p_z^2+M_f^2)^{1/2},
\end{equation}
where $n~(=0,1,2,...)$ represents the $n$th Landau level~(LL). 

To deal with the divergence in the vaccum part $\Omega_{f}^{\mathrm{0}}$, we take a smooth cutoff regularization procedure introduced in~\cite{Fukushima:2010fe}. A form factor $f_{\Lambda}(p_f)$ multiplying the integral kernel of $\Omega_{f}^{\mathrm{0}}$ is taken to avoid cutoff artifact, thus we have
 \begin{eqnarray}
\begin{aligned} \Omega_{f}^{\mathrm{0}}=-N_{c}\frac{\left|q_{f}\right| e B}{2 \pi} \sum_{n=0}^{\infty} \alpha_{n} \int_{-\infty}^{\infty} \frac{d p_{z}}{2 \pi} f_{\Lambda}^2\left(p_{f}\right)E_{f,n}\end{aligned},
\end{eqnarray}
where
 \begin{eqnarray}
f_{\Lambda}(p)=\sqrt{\frac{\Lambda^{2 N}}{\Lambda^{2 N}+p^{2 N}}},
\end{eqnarray}
$N=10$ is chosen in the numerical calculation. 
 One can see that $ f_{\Lambda}(p)$ is reduced to the sharp cutoff function $\theta(\Lambda-|\boldsymbol{p}|)$ in the $ N \rightarrow \infty $  limit.  Since the thermal part of $\Omega_f^T$ is not divergent, it is unnecessary to introduce a regularization function.

In fact, the original PNJL model with a background magnetic field can not describe well the IMC effect derived in LQCD. To solve this problem, a magnetic field dependent coupling constant for four-fermion interaction is proposed in Ref.~\cite{Ferreira18}. The coupling takes the form
\begin{equation}\label{}
G(B)=G_0\frac{1+a\zeta^2+b\zeta^3}{1+c\zeta^2+d\zeta^4},
\end{equation}
where $\zeta=\frac{eB}{\Lambda_{QCD}^2}$, with $\Lambda_{QCD}=300$ MeV. The parameters are $a = 0.108805$, $b =-1.0133\times10^{-4}$, $c= 0.02228$, and $d = 1.84558\times10^{-4}$. We will take such a  magnetic field dependent coupling in the present study.

 Other thermodynamic quantities relevant to the bulk properties of magnetized quark matter can be obtained from $\Omega$. 
 We take the model parameters obtained in~\cite{Rehberg96}:
$\Lambda=603.2$ MeV, $G_0\Lambda^{2}=1.835$, $K\Lambda^{5}=12.36$,
$m_{u,d}=5.5$  and $m_{s}=140.7$ MeV, determined
by fitting $f_{\pi}=92.4$ MeV,  $M_{\pi}=135.0$ MeV, $m_{K}=497.7$ MeV and $m_{\eta}=957.8$ MeV. $\mu_u=\mu_d=\mu_s$ is taken in the calculation.

\section{Numerical results and discussions }
In this section, we demonstrate the numerical results of the deformed QCD phase diagram in the presence of an external magnetic field, and discuss its relation with Landau quantization. We mainly concentrate on the first-order transition region at finite temperature.

\subsection{Deformed $\rho_{_B}-\mu_{_B}$ curves under magnetic field }

\begin{figure}[htbp]
\begin{center}
\includegraphics[scale=0.47]{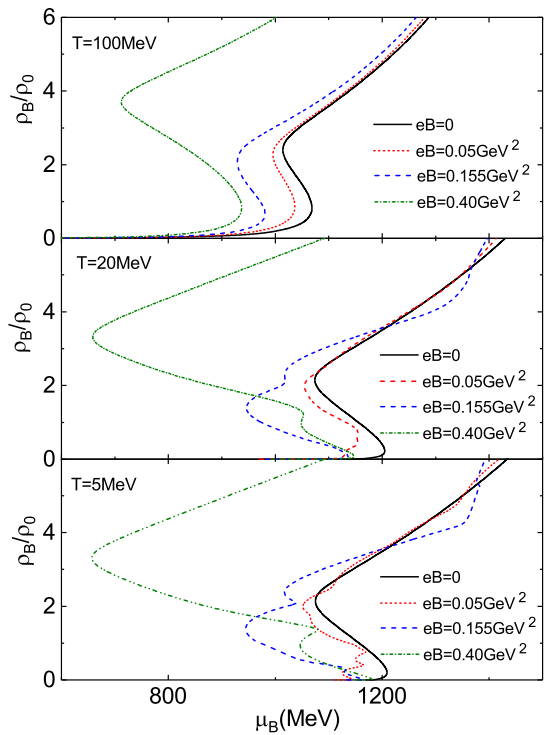}
\caption{\label{fig:figure1.eps}(color online) $\rho_{\scriptscriptstyle B}-\mu_{\scriptscriptstyle B}$ curves with $eB=0, 0.05, 0.155$, and $0.4\,\mathrm{GeV^2}$ at three different temperatures.  }
\end{center}
\end{figure}

To illustrate how the magnetic field strength affect the first-order transition, we present, in Fig.~\ref{fig:figure1.eps}, the $\rho_{_B}-\mu_{_B}$ relations at different temperatures with $eB=0, 0.05, 0.155$, and $0.4\,\mathrm{GeV^2}$, respectively. The upper panel shows that, at $T=100\,$MeV, the $\rho_{_B}-\mu_{_B}$ curve for each $eB$ has one single spinodal structure, which is a typical characteristic of a first-order transition. But with the decrease of temperature,  the $\rho_{_B}-\mu_{_B}$ curves become more and more complicated, as shown in the middle and lower panels. Particularly, multiple inflections appear for relatively smaller magnetic field at low temperatures.

The complicated twist in the $\rho_{_B}-\mu_{_B}$ curves at low temperatures are closely related to the Landau quantization. The filling of multiple Landau levels should be responsible for the number and locations where the inflection points appear in the $\rho_{_B}-\mu_{_B}$ curve.  To see this clearly, we start analysis from the quark number density. For each quark flavor $f$, the  number density can be derived as
 \begin{equation}\label{}
\rho_f=\frac{\partial\Omega}{\partial\mu_f}=\sum_{n=0}^{\infty}\rho_{f,n},
\end{equation}
where $\rho_{f,n}$ is the number density of the $n$th Landau level. Specifically,
 \begin{equation}\label{rhoin}
\rho_{f,n}\!\!=3\frac{q_fB}{2\pi}\alpha_n\!\!\int_{\!-\infty}^{\!\infty}\!\!\frac{\mathrm{d}p_z}{2\pi}\big[f_{f,n}^{+}-f_{f,n}^{-}\big],
\end{equation}
where
 \begin{equation}\label{}
 \!\!f_{f,n}^{+}\!\!=\!\!\frac{\Phi\,e^{-\frac{E_{f,n}-\mu_f}{T}}+2\bar\Phi\,e^{-2\frac{E_{f,n}-\mu_f}{T}}+\,e^{-3\frac{E_{f,n}-\mu_f}{T}}}{(1\!
+\!3\Phi e\!^{-\frac{E_{f,n}-\mu_f}{T}}\!\!+\!3\bar{\Phi} e^{-2\frac{E_{f,n}-\mu_f}{T}}\!\!+\!e^{-3\frac{E_{f,n}-\mu_f}{T}})},
 \end{equation}
  \begin{equation}\label{}
 \!\!f_{f,n}^{-}\!\!=\!\!\frac{\bar\Phi\,e^{-\frac{E_{f,n}+\mu_f}{T}}+2\Phi\,e^{-2\frac{E_{f,n}+\mu_f}{T}}+\,e^{-3\frac{E_{f,n}+\mu_f}{T}}}{(1\!
+\!3\Phi e\!^{-\frac{E_{f,n}+\mu_f}{T}}\!\!+\!3\bar{\Phi} e^{-2\frac{E_{f,n}+\mu_f}{T}}\!\!+\!e^{-3\frac{E_{f,n}+\mu_f}{T}})}.
 \end{equation}

\begin{figure}[htbp]
\begin{center}
\includegraphics[scale=0.42]{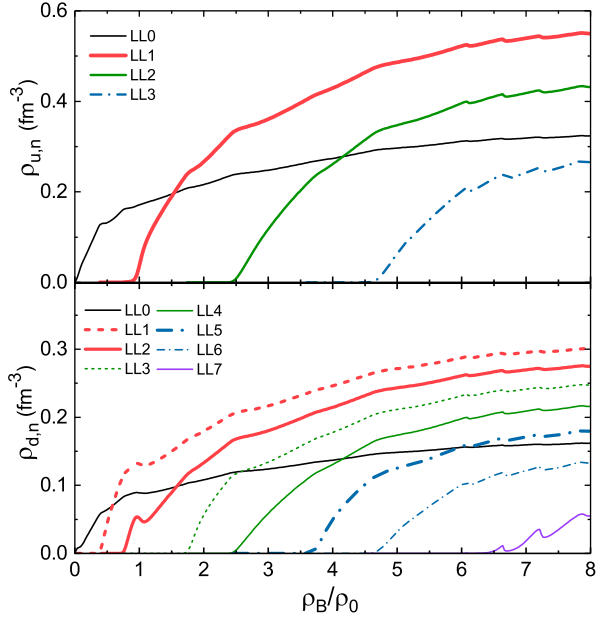}
\caption{\label{fig:figure2.eps}(color online) Curves of $\rho_{f,n}$ for the first few Landau levels as functions of $\rho_{\scriptscriptstyle B}$.  }
\end{center}
\end{figure}

 In the following, we will take  $T=5\,$MeV and $eB=0.05\,\mathrm{GeV^2}$ as an example to discuss the relation between  $\rho_{_B}-\mu_{_B}$ curve and the filling of Landau levels. The curve of $\rho_{f,n}$ for each Landau level as a function of $\rho_{\scriptscriptstyle B}$ is plotted in Fig.~\ref{fig:figure2.eps}. In this and all subsequent figures, LL0  means the lowest Landau level, LL1 is the first Landau level, LL2 is the second, and so forth. The upper and lower panels in Fig.~\ref{fig:figure2.eps} respectively illustrate  the number density of different Landau levels of up and down quarks.
In the density region of $0\!<\!\rho_{\scriptscriptstyle\!B}\!\leq\!8\rho_{0}$, the lowest four Landau levels for $u$ quarks are sequentially occupied as the baryon density increases, and the lowest eight Landau levels  are occupied for $d$ quarks. 

\begin{figure}[htbp]
\begin{center}
\includegraphics[scale=0.42]{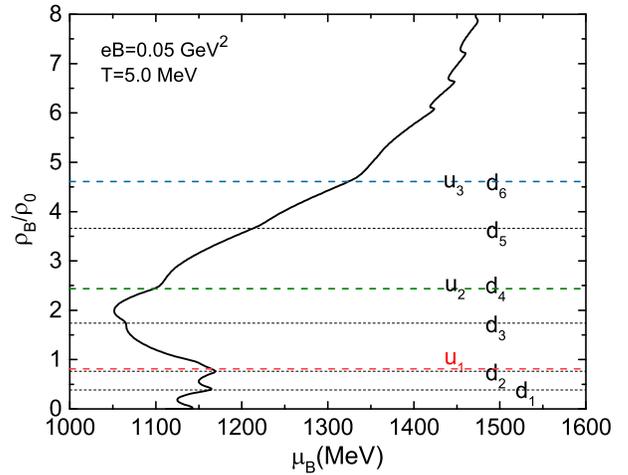}
\caption{\label{fig:figure3.eps}(color online)  $\rho_{\scriptscriptstyle\!B}-\mu_{\scriptscriptstyle B}$ curve for $eB=0.05\,\mathrm{GeV^2}$ at $T=5\,$MeV. The  subscript $i$ of $u_i$ and $d_i$ means the $i$th Landau level of $u$~(d) quark. A horizontal line marked with two Landau levels, such as $u_2$ and $d_4$, means that the two   Landau levels are filled (almost) at the same baryon density. }
\end{center}
\end{figure}

The corresponding  $\rho_{\scriptscriptstyle\!B}-\mu_{\scriptscriptstyle B}$ curve is demonstrated in Fig.~\ref{fig:figure3.eps}.
The thresholds of  different Landau levels are marked with the horizontal lines. The  subscript $n$ of $u_n$ and $d_n$ in  Fig.~\ref{fig:figure3.eps} means the $n$th Landau level of $u$~(d) quark. A horizontal line marked with two Landau levels, such as $u_2$ and $d_4$, means that the two   Landau levels are filled (almost) at the same baryon density.
It can be seen that when quarks fill a new Landau level a twist will appear in the $\rho_{\scriptscriptstyle\!B}-\mu_{\scriptscriptstyle B}$ curve.
 The twist induced by the Landau level is more distinct at small baryon density. Besides, the zigzag at $\rho_{\scriptscriptstyle\!B}>6\rho_{\scriptscriptstyle 0}$ results from the strange quark filling the new Landau levels. Therefore, the multiple Landau levels are responsible for the twisted $\rho_{\scriptscriptstyle\!B}-\mu_{\scriptscriptstyle B}$ relations when the magnetic field is considered.

 \begin{figure}[htbp]
\begin{center}
\includegraphics[scale=0.42]{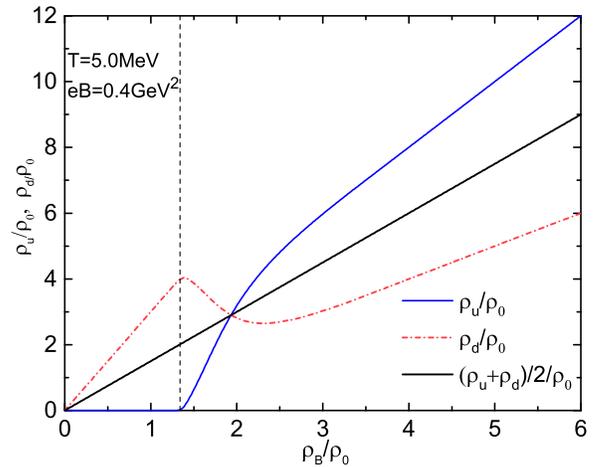}
\caption{\label{fig:figure4.eps}(color online)  $\rho_u$ and $\rho_d$ as functions of $\rho_{\scriptscriptstyle B}$. The fluctuation at low density is induced by $u$ quark filling the lowest Landau level LL0 and the fluctuation at high density is due to the strange quarks filling the new Landau level. }
\end{center}
\end{figure}

Since  $\Phi\approx\bar\Phi\approx0$ at low temperature, the contribution from gauge field  can be approximately neglected, so we have
   \begin{equation}\label{10302009} 
\rho_{f,n}\approx3\frac{q_f B}{2\pi^2}\alpha_n\!\int_{\!0}^{\!\infty}\!\!\frac{\mathrm{d}p_z}{(2\pi)}\bigg[\frac{1}{1+e^{3\frac{(E_{f,n}-\mu_f)}{T}}}\bigg].
 \end{equation}
 From Eq.~(\ref{10302009}), it is easy to know that the chemical potential of $\rho_{f,n}$ from zero to non-zero approximately satisfies the condition $\sqrt{M_f^2+2|q_f|Bn}=\mu_f$. For the lowest Landau level, the condition becomes $M_f=\mu_f$. For a large magnetic field,  the isospin symmetry is clearly broken since $q_u\neq\,q_d$.  For example, $M_u=431$ MeV and $M_d$=400 MeV are derived for  $eB=0.4~\mathrm{GeV^2}$  at zero density for T=5 MeV. Therefore the threshold of $\rho_{u,0}$ will be larger than that of  $\rho_{d,0}$, as shown in Fig.~\ref{fig:figure4.eps}.   This figure also shows that $\rho_{d,0}$ decreases 
 with the onset of $\rho_{u,0}$. Correspondingly, we can understand that the small zigzag at low density in the $\rho_B-\mu_B$ curve of $eB=0.4~\mathrm{GeV^2}$ in the lower panel of Fig.~\ref{fig:figure1.eps} is induced by that $u$ quark begins to occupy the lowest Landau energy~(LL0).

Eq.~(\ref{10302009}) also indicates that the maximum $n$ of the filled Landau level satisfies $n_{max}=$Floor$(\frac{\mu_f^2-M_f^2}{2|q_f|B})$. It is inversely proportional to the quark charge $q_f$ and the magnetic field strength $eB$, which is also indicated by the numerical results in Figs.~\ref{fig:figure1.eps} and~\ref{fig:figure2.eps}. At high density~(large chemical potential), the dynamic quark mass $M_f$ approaches the current quark mass after chiral symmetry restoration.
Since a $u$ quark is charged 2/3 and a $d$ quark is charged $-1/3$, when $u$ quarks fill one Landau level $d$ quarks will fill two.

\subsection{Entropy oscillation with the increase of density  }

Many studies have involved the de Haas-van Alphen effect of magnetized matter, a phenomenon related to the filling of  Landau levels, in which a physical quantity oscillates as a function of magnetic field intensity~\cite{Wang:2017pje,Aoki:2015mqa,Lugones:2016ytl,Wen:2016atg}. In this subsection, we discuss the entropy  oscillation as a function of baryon number density.

\begin{figure}[htbp]
\begin{center}
\includegraphics[scale=0.42]{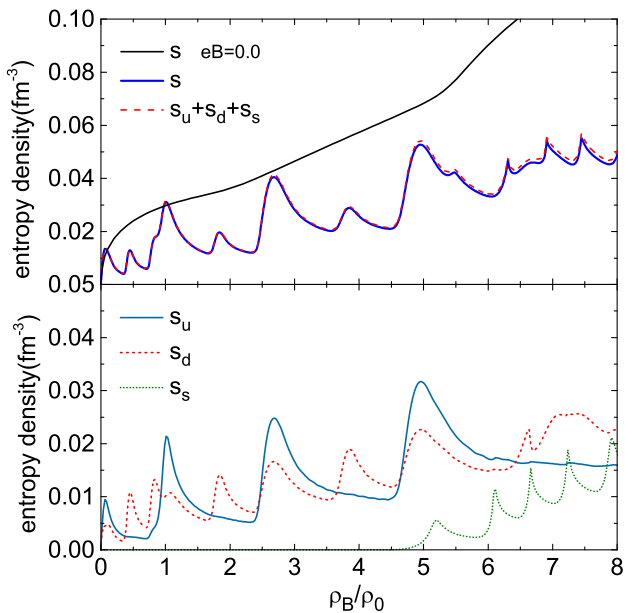}
\caption{\label{fig:figure5.eps}(color online) Entropy density without and with a magnetic field $eB=0.05~\mathrm{GeV^2}$ at T=5 MeV.   $s$ is the total entropy density; $s_i$ is the entropy density of quark flavor $i$.}
\end{center}
\end{figure}
The total entropy density $s$ as a function of baryon density is plotted in Fig.~\ref{fig:figure5.eps} without and with a magnetic field  $eB=0.05~\mathrm{GeV^2}$ at T=5 MeV.  A distinct oscillating behavior appears when the external magnetic field is considered. 
The numerical results in the upper panel show that the total entropy density $s$ is approximately equal to $s_u+s_d+s_s$. This can be understood since the contribution from the gauge sector is very small  at very low temperatures.

The lower panel of Fig.~\ref{fig:figure5.eps} describes the entropy densities of different quark flavor. It shows that the entropy density oscillation at low baryon densities mainly comes from the $u$ and $d$ quarks. The contribution of strange quarks  appear at high density.
It also shows that the number of the peaks of $s_d$ is almost twice of $s_u$. Since nearly half  of the peaks of $s_d$ appear at the same baryon densities with the peaks of $s_u$, the larger peaks of the total entropy density $s$ in the upper panel reflect the superposition of $s_u$ and $s_d$ before the strange quarks appear. The smaller peaks are only induced by $s_d$. Recalling the previous conclusion about the number of Landau levels filled by $u$ and $d$ quarks, it reminds us that the oscillation of entropy density may be also induced by quarks filling multiple Landau levels.

\begin{figure}[htbp]
\begin{center}
\includegraphics[scale=0.4]{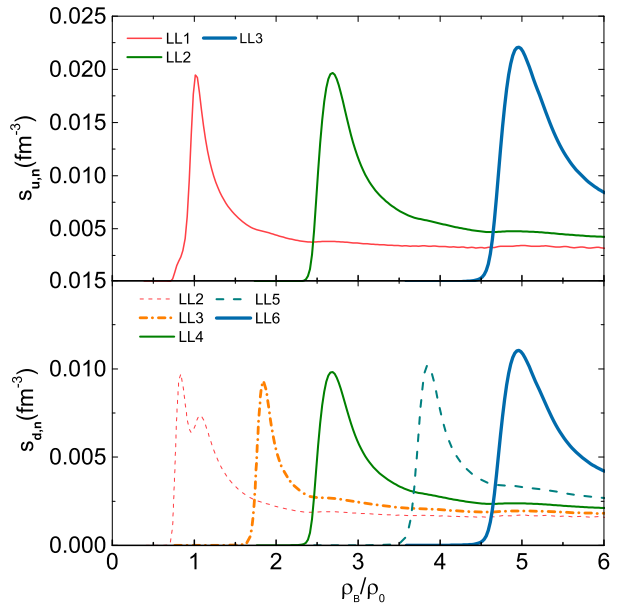}
\caption{\label{fig:figure6.eps}(color online) Entropy density $s_{f,n}$ of each Landau level as a function of baryon density. The upper~(lower) panel describes the entropy densities of  different Landau levels of $u$~($d$) quark.  }
\end{center}
\end{figure}

\begin{figure}[htbp]
\begin{center}
\includegraphics[scale=0.4]{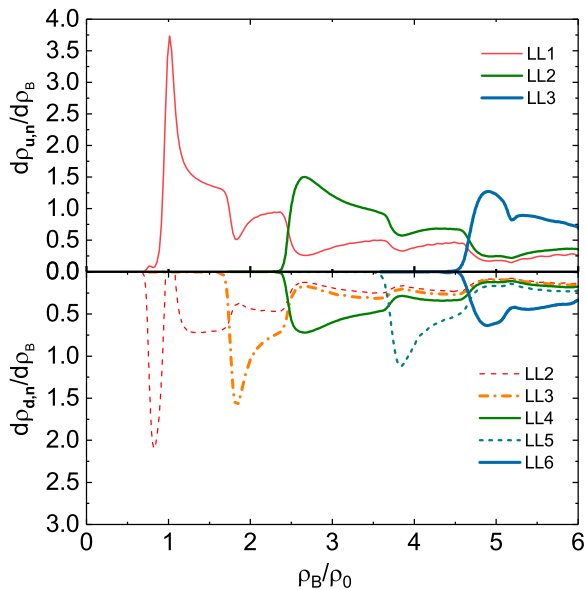}
\caption{\label{fig:figure7.eps}(color online) Differential of $\rho_{f,n}$ of the $n$th Landau level of quark flavor $f$ respect to $\rho_{\scriptscriptstyle B}$. }
\end{center}
\end{figure}

Considering the contribution of each Landau level to the total entropy density, we can decompose the total entropy density before the appearance of strange quarks as (contribution from gauge sector is neglected at low temperature)
     \begin{equation}\label{} 
 s\approx s_u+s_d=\sum_{f=u,d}\sum_{n=0}^{\infty}s_{f,n}.
 \end{equation} 
Fig.~\ref{fig:figure6.eps} illustrates the curves of $s_{f,n}$ with the increase of baryon density. It shows that each $s_{f,n}$ varies non-monotonically as the density increases.  The location of the peak of each $s_{f,n}$ corresponds to the density where $\text{d}\rho_{f,n} / \text{d}\rho_{_B}$ takes the maximum value. This can be seen by comparing Figs.~\ref{fig:figure6.eps}  with \ref{fig:figure7.eps}.

\begin{figure}[htbp]
\begin{center}
\includegraphics[scale=0.4]{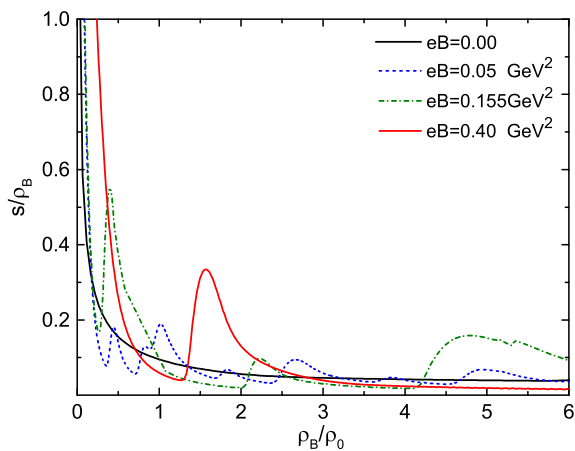}
\caption{\label{fig:figure8.eps}(color online)  Entropy per baryon  as functions of $\rho_{\scriptscriptstyle B}$ for $eB=0, 0.05, 0.155$, and $0.4\, \mathrm{GeV^2}$. }
\end{center}
\end{figure}

Fig.~\ref{fig:figure7.eps} describes the derivative of $\rho_{f,n}$ respect to $\rho_{_B}$. It shows  that $\text{d}\rho_{f,n} / \text{d}\rho_{_B}$ has also several minima at the locations where $\text{d}\rho_{u,n^{'}} / \text{d}\rho_{_B}$ or  $\text{d}\rho_{d,n^{'}} / \text{d}\rho_{_B}$  of the subsequent Landau levels take maxima. 
This can be understood from the following relation (before strange quarks appear at high density)
     \begin{equation}\label{33} 
\rho_{\scriptscriptstyle B}=\frac{1}{3}\sum_{f=u,d}\sum_{n=0}^{\infty}\rho_{f,n}.
 \end{equation}
For a fixed $\rho_{\scriptscriptstyle B}$,  any two $\text{d}\rho_{f,n} / \text{d}\rho_{_B}$ with different $f$ or $n$ are in a competitive relationship. The growth of one side must be accompanied by the decrease of the other side.

 In Fig.~\ref{fig:figure8.eps}, we present the entropy per baryon~($s/\rho_{_B}$) as functions of density for $eB=0, 0.05, 0.155$, and $0.4\, \mathrm{GeV^2}$, respectively.  It can be seen that, for a smaller magnetic field such as $eB=0.05\,\mathrm{GeV^2}$, the frequent oscillations occur  because more Landau levels are filled. The oscillations take place around the curve of $s/\rho_{_B}$ for $eB=0$.
The oscillation frequency decreases with the enhancement of  magnetic field strength. Furthermore,
the numerical analysis indicates that each valley in the $s/\rho_{_B}$ curves corresponds to the threshold of a new Landau level, and  each peak corresponds  to the global maximum of $\text{d}\rho_{f,n} / \text{d}\rho_{_B}$ for a Landau level.

\subsection{ Magnetic field dependence of QCD phase structure }
In this subsection, we analyze the magnetic field dependence of the QCD phase diagram, in particular the deformation of the first-order phase transition. 

\begin{figure*} [htbp]
\centering
\includegraphics[width=.9\linewidth]{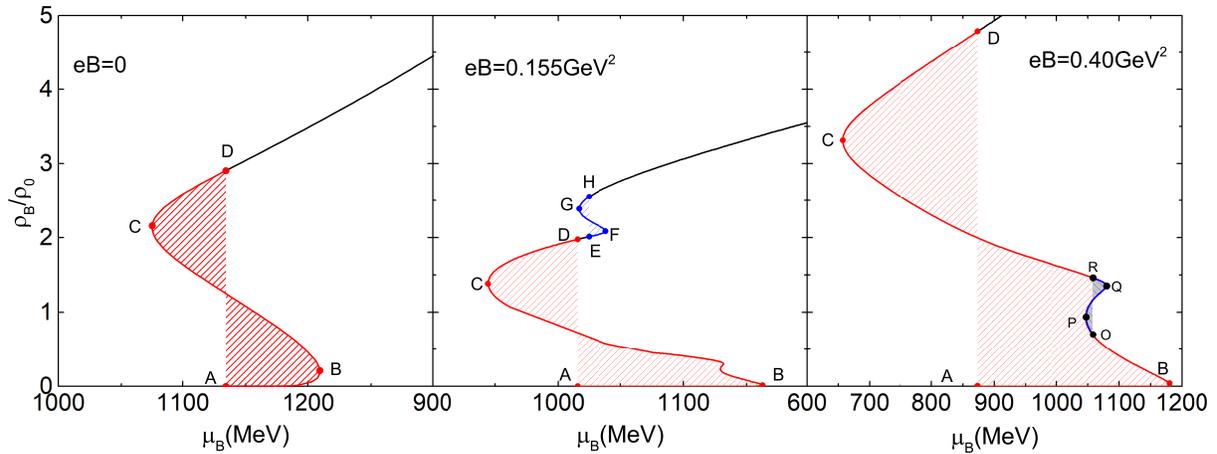}
\caption{(color online) $\rho_{_B}-\mu_{_B}$ curves without~($eB= 0$) and with an external magnetic field for $eB= 0.155$ and $0.4\,\mathrm{GeV^2}$ at $T=5\,$MeV.
For $eB=0$, the first-order transition takes place at the locations of $A$ and $D$. For $eB=0.155$, two first-order transitions occur at the locations of $A$ and $D$ as well as $E$ and $H$. For $eB=0.4$, the first-order transition takes place at the locations of $A$ and $D$. 
} 
\label{fig:figure9.eps}
\end{figure*}

We first discuss the first-order  transition under different magnetic field intensity for a fixed temperature $T=5\,$MeV.
The $\rho_{_B}-\mu_{_B}$ curves without~($eB=0$) and with an external magnetic field for $eB=0.155$ and $0.4\,\mathrm{GeV^2}$  are plotted in Fig.~\ref{fig:figure9.eps}.
For the case of zero magnetic field, the first-order transition occurs at $\rho_{ _A}$ and $\rho_{_D}$.  The two regions of $A$-$B$ and $C$-$D$ are the metastable phases,  in which nucleation and bubble formation possibly occur. The region of $B$-$C$ is the mechanically unstable phase because of $\partial p/\partial\rho<0$, which is known as the spinodal region. When the bulk uniform matter enters into this region, a small fluctuation in density will lead to phase separation via the spinodal decomposition. Generally, for an equilibrium transition, the unstable phase can not be observed. But it is difficult to estimate the role it plays on observables such as the particle fluctuations in a rapid expanding system.

For $eB=0.155\,\mathrm{GeV^2}$, the phase structure is quite different with that of zero magnetic field. Two first-order phase transitions, from $A$ to $D$ and from $E$ to $H$, take place. The locations of the two transitions are determined according to the conditions for phase equilibrium: $T_A=T_D,\mu_A=\mu_D$ and $P_A=P_D$ as well as $T_E=T_H,\mu_E=\mu_H$ and $P_E=P_H$.
Moreover, between the two first-order transitions, a stable phase of the magnetized matter exists in the region of $D$-$E$. Such a special phase structure is driven by the Landau quantization with the filling of different Landau levels by quarks. 

With the increase of magnetic field intensity, for example $eB=0.4\,\mathrm{GeV^2}$, the right panel of Fig.~\ref{fig:figure9.eps} shows that the first-order transition occur at $\rho_{ _A}$ and $\rho_{_D}$, similar with the case of $eB=0$. However, there exists a region marked as $O$-$P$-$Q$-$R$, which has a $\rho_{_B}-\mu_{_B}$ structure opposite to a standard first-order transition. At the locations $O$ and $R$, the  conditions for two phase equilibrium are fulfilled, but it is not a first-order transition because $O$ and $R$ lie in the unstable phase. On the other hand, bulk matter in the interval of $P$-$Q$ are metastable, which can not be observed for an equilibrium transition, since the phase transition from  $\rho_{ _D}$ to $\rho_{_A}$ will first take place.

\begin{figure}[htbp]
\begin{center}
\includegraphics[scale=0.35]{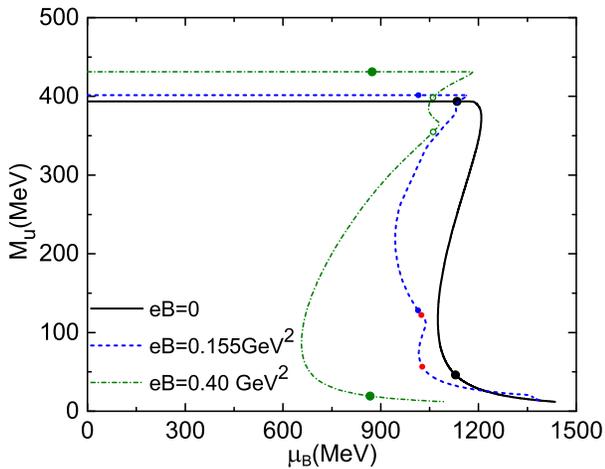}
\caption{\label{fig:figure10.eps}(color online) Dynamical quark mass of $u$ quark as a function of baryon chemical potential for $eB=0, 0.155$ and $0.4\,\mathrm{GeV^2}$, respectively, at T=5 MeV. }
\end{center}
\end{figure}

The dynamical mass of $u$ quark as functions of baryon chemical potential are illustrated in Fig.~\ref{fig:figure10.eps}  for $eB=0, 0.155$ and $0.4\,\mathrm{GeV^2}$.  It can be seen that $M_u$ in the chiral breaking phase increases with the enhancement of magnetic field intensity, which reflects the magnetic catalysis  effect at low temperature. It is consistent with LQCD calculation \cite{Buividovich10}. The solid dots with the same color on $M_u-\mu_{_B}$ curves indicate the locations where the first-order transition takes place. The circles on the curve of  $eB=0.4\,\mathrm{GeV^2}$ does not mean a first-order transition although the conditions for phase equilibrium are fulfilled. The curves of   $eB=0.155\,\mathrm{GeV^2}$ indicates that two first-order transition can take place. Similar phenomenon was discovered at zero temperature~\cite{ Allen:2013lda, Denke13, Grunfeld:2014qfa}.

\begin{figure*} [hbtp]
\centering
\includegraphics[width=.9\linewidth]{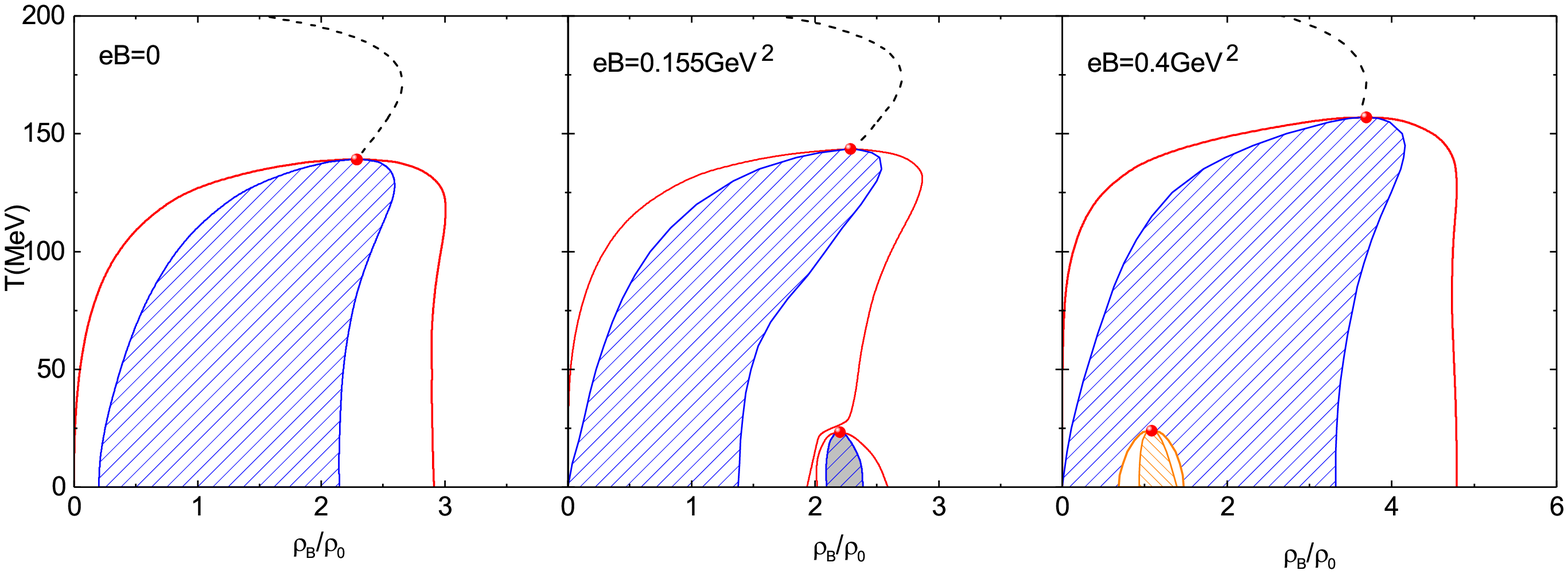}
\caption{(color online) Phase diagrams in the $T-\rho_{_B}$ plane for $eB=0, 0.155$ and $0.4\,\mathrm{GeV^2}$. The dashed line in each panel corresponds to the chiral crossover transition at high temperatures. The first-order transitions are marked with the solid red lines. The blue lines indicate the spinodal regions. The orange lines and region for $eB=0.4\,\mathrm{GeV^2}$ indicate the range with a structure opposite to a standard first-order transition.  } 
\label{fig:figure11.eps}
\end{figure*}

The complete phase diagram of the chiral phase transition is demonstrated in Fig.~\ref{fig:figure11.eps}. For the first-order transition at low temperatures, the associated metastable and unstable regions are also included.   This figure distinctly illustrates the deformed QCD phase structures driven by magnetic field with different  field intensity. When the external magnetic field is larger than $eB=0.4\,\mathrm{GeV^2}$, the magnetized quark matter has a similar phase structure to $eB=0.4\,\mathrm{GeV^2}$. Two first-order transitions exist in the vicinity of $eB=0.155\,\mathrm{GeV^2}$. For a smaller magnetic field such as $eB=0.05\,\mathrm{GeV^2}$, more Landau levels will be occupied and the corresponding phase diagram is more complicated, as shown in Fig.~\ref{fig:figure3.eps}. 

\begin{figure*} [htbp]
\centering
\includegraphics[width=.9\linewidth]{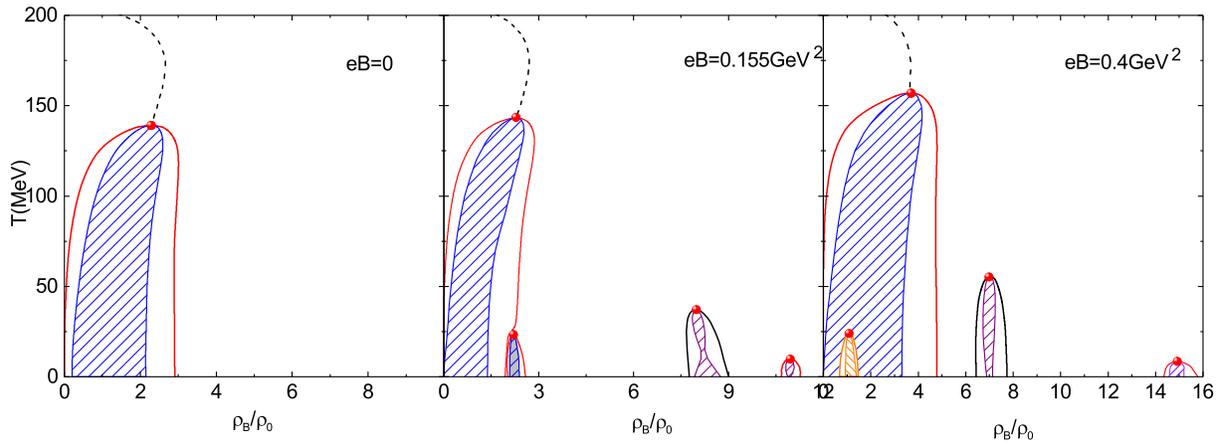}
\caption{(color online) Phase diagrams in the $T-\rho_{_B}$ plane including the contribution of strange quarks at high density for $eB=0, 0.155$ and $0.4\,\mathrm{GeV^2}$.
The first-order transitions of strange quark are marked with the solid black lines. The purple lines indicate the corresponding spinodal regions} 
\label{fig:figure12.eps}
\end{figure*}

\begin{figure*} [htbp]
\centering
\includegraphics[width=.9\linewidth]{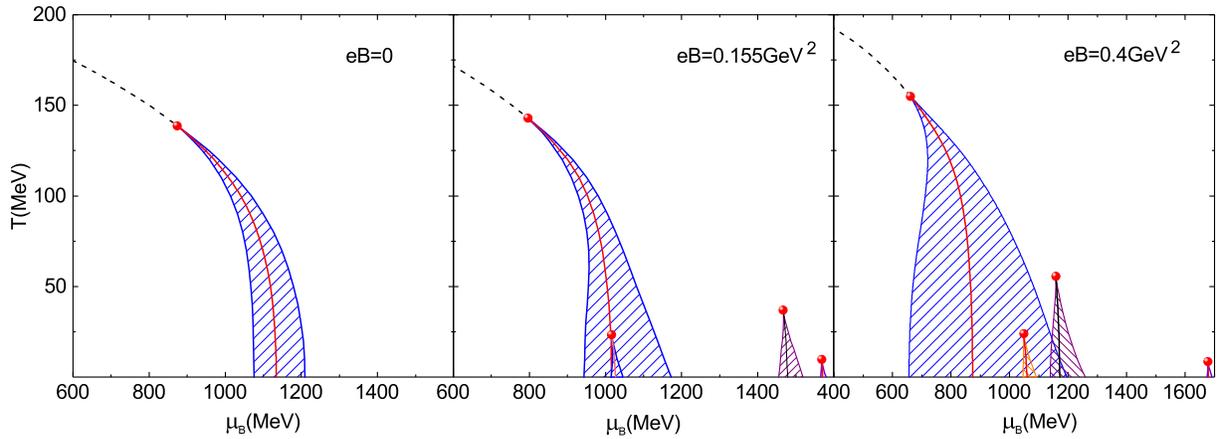}
\caption{(color online) Phase diagrams in the $T-\mu_{_B}$ plane including the contribution of strange quarks at large chemical potential for $eB=0, 0.155$ and $0.4\,\mathrm{GeV^2}$. } 
\label{fig:figure13.eps}
\end{figure*}

The phase diagrams including the contribution of strange quark at high density~(large chemical potential) are illustrated in Figs.~\ref{fig:figure12.eps} and \ref{fig:figure13.eps}. Compared with the phase structure without an external magnetic field, the two figures indicate that the first-order transitions in the strange quark sector can also be induced by Landau quantization when $s$ quarks fill the different Landau levels. The locations where the first-order transitions take place depend on the magnetic field intensity. One can also refer to \cite{Ferreira18} for the related discussion about the first-order transition in strange quark sector.

 In this study, we find the phase structure of strongly interacting matter under an external magnetic field highly depends on the field intensity. Experimently, the high-density region
can possibly be reached in future heavy-ion collisions at RHIC, NICA and FAIR.
 At the same time, the magnetic field with $eB\lesssim0.155\,\mathrm{GeV^2}$ can possibly be created in the non-central collisions, therefore, the multi-first-order transitions or more complicated phase structure of light quarks may give birth to some observable effects in the beam energy scan with relatively lower collision energies.

\section{summary}
In this study, we investigated the chiral phase transition in the presence of an external magnetic field in the improved PNJL model. The calculations show that the phase structure of magnetized quark matter strongly depends on the intensity of magnetic field. Different from the first-order phase transition without a magnetic field, two first-order transitions or more complicated phase transition in the light quark sector can occur for $eB\lesssim0.155\,\mathrm{GeV^2}$. The study also indicates that the deformation of the phase structure under an external magnetic field is attributed to the Landau quantization with the filling of different Landau levels.  Generally, for a relatively smaller magnetic field, more Landau levels will be filled which leads to a twisted $\rho{_B}-\mu_{B}$ relation and then produce a complicated phase structure. 

We also found that the distribution of quarks at multiple Landau levels causes the entropy density oscillation due to the Landau quantization.  The numerical results indicate that the entropy density as well as the entropy per baryon begin to increase at the threshold of a new Landau level. Each peak of the entropy density~(entropy per baryon) corresponds to a maximum value of $\partial\rho_{f,n} / \partial\rho_{_B}$ of a Landau level.

If the high-density quark matter and $eB\sim0.155\,\mathrm{GeV^2}$ could be created in the non-central heavy-ion collisions, some signals different from the standard first-order transition may manifest in future experiments. However, it is difficult in measurements due to the decay of magnetic field in the expansion. More simulations are needed to catch the relevant signatures.  This study is also referential to investigate the magnetized neutron star matter with a quark core or a magnetized quark star.

\begin{acknowledgments}
This work is supported by the National Natural Science Foundation of China under
Grant No. 11875213 and the
Natural Science Basic Research Plan in Shaanxi Province of
China (Program No. 2019JM-050).
\end{acknowledgments}



\begin{thebibliography}{}
\bibitem{Gupta11} S. Gupta, X. F. Luo, B. Mohanty, H. G. Ritter, and N. Xu, Science  {\bf 332},  1525 (2011).
\bibitem{Aoki06} Y. Aoki, G. Endrodi, Z. Fodor, S. D. Katz, K. K. Szabo, Nature    {\bf 443}, 675 (2006).
\bibitem{Bazavov12} A. Bazavov, et al., hotQCD Collaboration, Phys. Rev. D  {\bf 85},  054503 (2012).
\bibitem{Borsanyi13} S. Bors\'anyi, Z. Fodor, S. D. Katz, S. Krieg, C. Ratti
and K. K. Szab\'o, Phys. Rev. Lett. {\bf 111},  062005 (2013).
\bibitem{Bazavov14} A. Bazavov,  et al., hotQCD Collaboration, Phys. Rev. D. {\bf 90}, 094503  (2014). 
\bibitem{Bazavov17} A. Bazavov,  et al., hotQCD Collaboration, Phys. Rev. D {\bf 96}, 074510 (2017).
\bibitem{Borsanyi14} S. Bors\'anyi, Z. Fodor, C. Hoelbling, S. D. Katz, S. Krieg,
and K. K. Sabz\'o, Phys. Lett. B {\bf 730},  99 (2014).

\bibitem{Fukushima04}  K. Fukushima, Phys. Lett. B  {\bf 591}, (2004)  277; Phys. Rev. D {\bf 77}, 114028 (2008).
\bibitem{Ratti06} C. Ratti, M. A. Thaler, and W. Weise, Phys. Rev. D {\bf 73},  014019 (2006).
\bibitem{Costa10}  P. Costa, M. C. Ruivo, C. A. de Sousa, and H. Hansen, Symmetry {\bf 2}, 1338 (2010).
\bibitem{Fu08} W. J. Fu, Z. Zhang, and Y. X. Liu, Phys. Rev. D {\bf 77},  014006  (2008).
\bibitem{Sasaki12} T. Sasaki, J. Takahashi, Y. Sakai,  H. Kouno, and M. Yahiro, Phys. Rev. D {\bf 85},  056009 (2012).
\bibitem{Schaefer10} B. J. Schaefer, M. Wagner, and J. Wambach, Phys. Rev. D {\bf 81}, 074013 (2010).
\bibitem{Skokov11}  V. Skokov, B. Friman, and K. Redlich, Phys. Rev. C {\bf 83}, 054904  (2011).

\bibitem{Qin11} S. X. Qin,  L. Chang, H. Chen, Y. X. Liu, and C. D. Roberts, Phys. Rev. Lett. {\bf 106}, 172301 (2011). 
\bibitem{Gao16} F. Gao, J. Chen, Y. X. Liu, S. X. Qin, C. D. Roberts, and S. M. Schmidt, Phys. Rev. D {\bf 93},   094019 (2016).
\bibitem{Fischer14} C. S. Fischer, J. Luecker, and C. A. Welzbacher. Phys. Rev. D {\bf 90},  034022 (2014).
\bibitem{Shi14} C. Shi, Y. L. Wang, Y. Jiang, Z. F. Cui, H. S. Zong, JHEP {\bf 1407},  014 (2014). 

\bibitem{Aggarwal10} M. M. Aggarwal,  et al., STAR Collaboration, Phys. Rev. Lett. {\bf 105},  022302  (2010).
\bibitem{Adamczyk14}L. Adamczyk,  et al., STAR Collaboration, Phys. Rev. Lett. {\bf 112 }, 032302 (2014).
\bibitem{Luo2014} X. Luo (for the STAR Collaboration), PoS(CPOD2014) (2015) 019.
\bibitem{Luo2016} X. Luo, Nucl. Phys. A  {\bf 956}, 75  (2016). 
\bibitem{Luo2017} X. Luo and N. Xu, Nucl. Sci. Tech. {\bf 28}, 112 (2017).


\bibitem{Kharzeev13} D. Kharzeev, K. Landsteiner, A. Schmitt, and Ho-Ung Yee, Lect. Notes Phys. {\bf 971}, 1 (2013).

\bibitem{Miransky15} V. A. Miransky and I. A. Shovkovy, Phys. Rept. {\bf 576}, 1 (2015).
\bibitem{Aadersen16} J. O. Aadersen and W. R. Naylor, A. Tranberg,Rev. Mod. Phys. {\bf 88}, 025001, (2016).



\bibitem{Bocquet95} M. Bocquet, S. Bonazzola, E. Gourgoulhon, and J. Novak, Astron. Astrophys. {\bf 301}, 757 (1995).
\bibitem{Ferrer10} E. J. Ferrer, V. de la Incera, J. P. Keith, I. Portillo, P. L. Springsteen, Phys. Rev. C {\bf 82}, 065802 (2010).



\bibitem{Kharzeev08} D. E. Kharzeev, L. D. McLerran, and H. J. Warringa, Nucl. Phys. A {\bf 803}, 227 (2008).
\bibitem{Skokov09} V. Skokov, A. Y. Illarionov, and V. Toneev, Int. J. Mod. Phys. A {\bf 24}, 5925 (2009).

\bibitem{Voronyuk11} V. Voronyuk, V. D. Toneev, W. Cassing, E. L. Bratkovskaya, V. P. Konchakovski, and S. A. Voloshin, Phys. Rev. C {\bf 83}, 054911 (2011)
\bibitem{Bzdak12} A. Bzdak, V. Skokov, Phys. Lett. B {\bf 710}, 171 (2012)
\bibitem{Deng12} W. T. Deng, X. G. Huang, Phys. Rev. C {\bf 85}, 044907 (2012)




\bibitem{Buividovich09} P.~V.~Buividovich, M.~N.~Chernodub, E. V. Luschevskaya, and M. I. Polikarpov, Phys. Rev. D {\bf 80}, 054503 (2009).


\bibitem{Kharzeev11} D. E. Kharzeev and  H. U. Yee,  Phys. Rev. D {\bf 83}, 085007 (2011).
\bibitem{Metlitski05} M. A. Metlitski and A. R. Zhitnitsky, Phys. Rev. D {\bf 72}, 045011 (2005).


\bibitem{Burnier11} Y. Burnier, D. E. Kharzeev, J. Liao, and H. U. Yee,  Phys. Rev. Lett. {\bf 107}, 052303 (2011).
 \bibitem{Gorbar11} E. V. Gorbar, V. A. Miransky, I. A. Shovkovy, Phys. Rev. D {\bf 83}, 085003 (2011).
\bibitem{Fukushima08} K. Fukushima, D. E. Kharzeev, H. J. Warringa, Phys. Rev. D {\bf 78}, 074033 (2008). 
\bibitem{Fukushima10} K. Fukushima, M. Ruggieri, Phys. Rev. D {\bf 82}, 054001 (2010).
\bibitem{Gatto12} R. Gatto, M. Ruggieri, Phys. Rev. D {\bf 85}, 054013 (2012).
\bibitem{Ruggieri11} M. Ruggieri, Phys. Rev. D {\bf 84}, 014011 (2011).
 \bibitem{Fu2013} W.~J.~Fu, Phys.\ Rev.\ D {\bf 88}, 014009 (2013).%
\bibitem{Bayona11} C. A. B. Bayona, K. Peeters, and M. Zamaklar, J. High Energy Phys. {\bf 1106}, 092 (2011).



\bibitem{Buividovich10} P. Buividovich, M. N. Chernodub, E. V. Luschevskaya, and M. I. Polikarpov, Phys. Lett. B {\bf 682} 484 (2010).

\bibitem{elia}M. D'Elia, S. Mukherjee, F. Sanfilippo, Phys. Rev. D {\bf 82}, 051501(R)  (2010) .
\bibitem{Ilgenfritz12} E. M. Ilgenfritz, M. Kalinowski, M. M\"{u}ller-Preussker, B. Petersson, A. Schreiber, Phys. Rev. D {\bf 85} (2012) 114504.


 \bibitem{Bali:2011qj} G.~S.~Bali, F.~Bruckmann, G.~Endrodi, Z.~Fodor, S.~D.~Katz, S.~Krieg, A.~Schafer, and K~K.~Szabo, JHEP {\bf 1202}, 044 (2012).
 \bibitem{Bali:2012zg} G.~S.~Bali, F.~Bruckmann, G.~Endrodi, Z.~Fodor, S.~D.~Katz and A.~Schafer, Phys.\ Rev.\ D {\bf 86}, 071502(R) (2012).
 

 
 
\bibitem{Fukushima12} K. Fukushima and J. M. Pawlowski, Phys. Rev. D {\bf 86}, 076013 (2012).
\bibitem{Chao13} J. Chao, P. Chu, and M. Huang, Phys. Rev. D {\bf 88}, 054009 (2013).
\bibitem{Fraga13}  E. S. Fraga, J. Noronha, and L. F. Palhares, Phys. Rev. D {\bf 87,} 114014 (2013).
\bibitem{Fukushima13} K. Fukushima and Y. Hidaka, Phys. Rev. Lett. {\bf 110}, 031601 (2013).
\bibitem{Fukushima16} K. Fukushima and Y. Hidaka, Phys. Rev. Lett. {\bf 117}, 102301 (2016).

\bibitem{Fayazbakhsh14} S. Fayazbakhsh, and N. Sadooghi, Phys. Rev. D {\bf 90}, 105030 (2014).
\bibitem{Ferreira142} M. Ferreira, P. Costa, and C. Provid\^encia, Phys. Rev. D {\bf 90}, 016012 (2014).

\bibitem{Costa15} P. Costa, M. Ferreira, D. P. Menezes, J. Moreira, and
C. Provid\^encia, Phys. Rev. D {\bf 92}, 036012 (2015).
\bibitem{Fraga14} E. S. Fraga, B. W. Mintz, and J. Schaffner-Bielich, Phys. Lett. B {\bf 731}, 154 (2014).
\bibitem{Yu15} L. Yu, J. Van Doorsselaere, and M. Huang, Phys. Rev. D {\bf 91}, 074011 (2015).

\bibitem{Ayala16} A. Ayala, C. A. Dominguez, L. A. Hernandez, M. Loewe, and R. Zamora, Phys. Lett. B {\bf 759}, 99 (2016).
\bibitem{Pagura17} V. P. Pagura, D. Gomez Dumm, S. Noguera, and N. N. Scoccola,  Phys. Rev. D {\bf 95}, 034013 (2017).

\bibitem{Menezes:2008qt} D.~P.~Menezes, M.~B. Pinto, S.~S.~Avancini, A.~P. Martinez, and C.~Provid\^encia, Phys.\ Rev.\ C {\bf 79}, 035807 (2009).  
\bibitem{Ferreira18} M. Ferreira, P. Costa, and C. Provid\^encia, Phys. Rev. D {\bf 97}, 014014 (2018).

\bibitem{Robner07} S. R\"{o}{\ss}ner, C. Ratti, and W. Weise, Phys. Rev. D {\bf 75},  034007 (2007).
\bibitem{Fukushima:2010fe} K.~Fukushima, M.~Ruggieri, and R.~Gatto, Phys.\ Rev.\ D {\bf 81}, 114031 (2010).

\bibitem{Rehberg96} P. Rehberg, S. P. Klevansky, and J. H\"ufner, Phys. Rev. C {\bf 53},  410 (1996).
 
\bibitem{Wen:2016atg} X.~J.~Wen and J.~J.~Liang, Phys.\ Rev.\ D {\bf 94}, 014005 (2016).

 \bibitem{Wang:2017pje} L.~Wang and G.~Cao, Phys.\ Rev.\ D {\bf 97}, 034014 (2018)
 \bibitem{Aoki:2015mqa} K.~I.~Aoki, H.~Uoi, and M.~Yamada, Phys.\ Lett.\ B {\bf 753}, 580 (2016)
 \bibitem{Lugones:2016ytl} G.~Lugones and A.~G.~Grunfeld, Phys.\ Rev.\ C {\bf 95}, 015804 (2017)

 \bibitem{Allen:2013lda} P.~G.~Allen and N.~N.~Scoccola, Phys.\ Rev.\ D {\bf 88}, 094005 (2013).%
\bibitem{Denke13} R. Z. Denke and M. B. Pinto, Phys. Rev. D {\bf 88}, 056008 (2013).
\bibitem{Grunfeld:2014qfa} A.~G.~Grunfeld, D.~P.~Menezes, M.~B.~Pinto, and N.~N.~Scoccola, Phys.\ Rev.\ D {\bf 90}, no. 4, 044024 (2014).






%
%
%
%
%
%
%
%
%
%
%
%
%
%
%
%
%
%
%
%
%
%
%
%
%
%
%
%
%
%
%
%
%
%
%
%



\end{thebibliography}
\end{document}